\documentclass[journal=jacsat,manuscript=communication,layout=twocolumn]{achemso}

\usepackage{graphicx} 

 \SectionsOn

\mciteErrorOnUnknownfalse

\usepackage{siunitx}

\DeclareSIUnit\Molar{\textsc{M}}
\DeclareSIUnit\bar{bar}

\usepackage{physics}
\usepackage{amsmath}
\usepackage{booktabs}
\usepackage{caption}

\usepackage[capitalize]{cleveref}

\usepackage{bbold}

\title{Radio-Frequency Sweeps at $\mu$T Fields for Parahydrogen-Induced Polarization of Biomolecules}

\author{Alastair Marshall}
\altaffiliation{Contributed equally to this work}
\affiliation{NVision Imaging Technologies GmbH, 89081 Ulm, Germany}
\alsoaffiliation{Institute for Quantum Optics (IQO) and Center for Integrated Quantum Science and Technology (IQST), Albert-Einstein-Allee 11, Universit{\"a}t Ulm, D-89081 Ulm, Germany}

\author{Alon Salhov}
\altaffiliation{Contributed equally to this work}
\affiliation{NVision Imaging Technologies GmbH, 89081 Ulm, Germany}
\alsoaffiliation[HUJI]
{Racah Institute of Physics, The Hebrew University of Jerusalem, Jerusalem, 91904, Givat Ram, Israel}
\author{Martin Gierse}
\alsoaffiliation{Institute for Quantum Optics (IQO) and Center for Integrated Quantum Science and Technology (IQST), Albert-Einstein-Allee 11, Universit{\"a}t Ulm, D-89081 Ulm, Germany}
\author{Christoph M\"uller}
\author{Michael Keim}
\author{Sebastian Lucas}
\author{Anna Parker}
\author{Jochen Scheuer}

\author{Christophoros Vassiliou}

\author{Philipp Neumann}
\author{Fedor Jelezko}
\alsoaffiliation{Institute for Quantum Optics (IQO) and Center for Integrated Quantum Science and Technology (IQST), Albert-Einstein-Allee 11, Universit{\"a}t Ulm, D-89081 Ulm, Germany}
\author{Alex Retzker}
\alsoaffiliation[HUJI]
{Racah Institute of Physics, The Hebrew University of Jerusalem, Jerusalem, 91904, Givat Ram, Israel}
\author{John W. Blanchard}
\author{Ilai Schwartz}
\author{Stephan Knecht}
\email{stephan@nvision-imaging.com}

\affiliation{NVision Imaging Technologies GmbH, 89081 Ulm, Germany}

\date{\today}

\begin{document}

\begin{abstract}
Magnetic resonance imaging of $^{13}$C-labeled metabolites enhanced by parahydrogen-induced polarization (PHIP) can enable real-time monitoring of processes within the body.
We introduce a robust, easily implementable technique for transferring parahydrogen-derived singlet order into $^{13}$C magnetization using adiabatic radio-frequency sweeps at $ \mu$T fields.
We experimentally demonstrate the applicability of this technique to several molecules, including some molecules relevant for metabolic imaging, where we show significant improvements in the achievable polarization, in some cases reaching above 60\%. 
Furthermore, we introduce a site-selective deuteration scheme, where deuterium is included in the coupling network of a pyruvate ester to enhance the efficiency of the polarization transfer.
These improvements are enabled by the fact that the transfer protocol avoids relaxation induced by strongly coupled quadrupolar nuclei.
\end{abstract}

\maketitle

\section{Introduction}
Magnetic resonance imaging (MRI) is a widely used medical imaging procedure, where anatomical images are obtained by manipulating and measuring the nuclear-spin magnetization arising from water and fat in human tissues\cite{Kurhanewicz2011, Golman2006}.
Over the past two decades, significant research effort has been invested toward the development of ${}^{13}$C spectroscopic imaging as a functional imaging technique to probe metabolic activity\cite{kurhanewicz2019hyperpolarized,wang2019hyperpolarized}.
To compensate for the significantly reduced sensitivity of ${}^{13}$C readout, the technique utilizes injectable ${}^{13}$C-labeled substrates with greatly enhanced ($>10^4$) nonequilibrium nuclear spin polarization \cite{haakeEfficientNMRPulse1996a}.
Following injection of this hyperpolarized solution, metabolic conversion of the substrate into other hyperpolarized molecules (distinguished by their different ${}^{13}$C chemical shifts) allows for mapping of metabolic activity.
This has been used for imaging both animals and humans\cite{Golman2006,harris2009kinetics,moreno2010competition}.

To date, most ${}^{13}$C metabolic imaging has relied on dynamic nuclear polarization (DNP), which involves microwave-driven transfer of electron spin polarization to nuclear spins \cite{Abragam1978, wenckebach2016essentials}.
While electron spin polarization is generally much higher than nuclear spin polarization at finite fields, achieving near-unity electron spin polarization requires that samples are cooled to cryogenic temperatures (ca. 1\,K) in high magnetic fields (typically several T).
Instrumentation for producing clinical volumes of highly polarized ($>10$\,\%) metabolites is therefore costly and technically complex.

Parahydrogen-induced polarization (PHIP) is an alternative spin-polarization method\cite{pravica1988net,bowers1987parahydrogen}, where hydrogen gas is prepared in its nuclear spin singlet state (i.e., enriched in the para spin isomer) and subsequently introduced into a molecule by hydrogenation of an unsaturated precursor.
Subsequently, the parahydrogen-derived spin order is converted into ${}^{13}$C magnetization for later use \cite{eillsPolarizationTransferField2019, Knecht2021,Reineri2015}. 
Significant effort has been spent to apply PHIP to hyperpolarize biomolecules, such as pyruvate and fumarate, suitable for analysis of human pathologies \cite{hovenerParahydrogenBasedHyperpolarizationBiomedicine2018,Reineri2015,Knecht2021}.

\begin{figure*}
    \centering
    \includegraphics[width=0.85\textwidth]{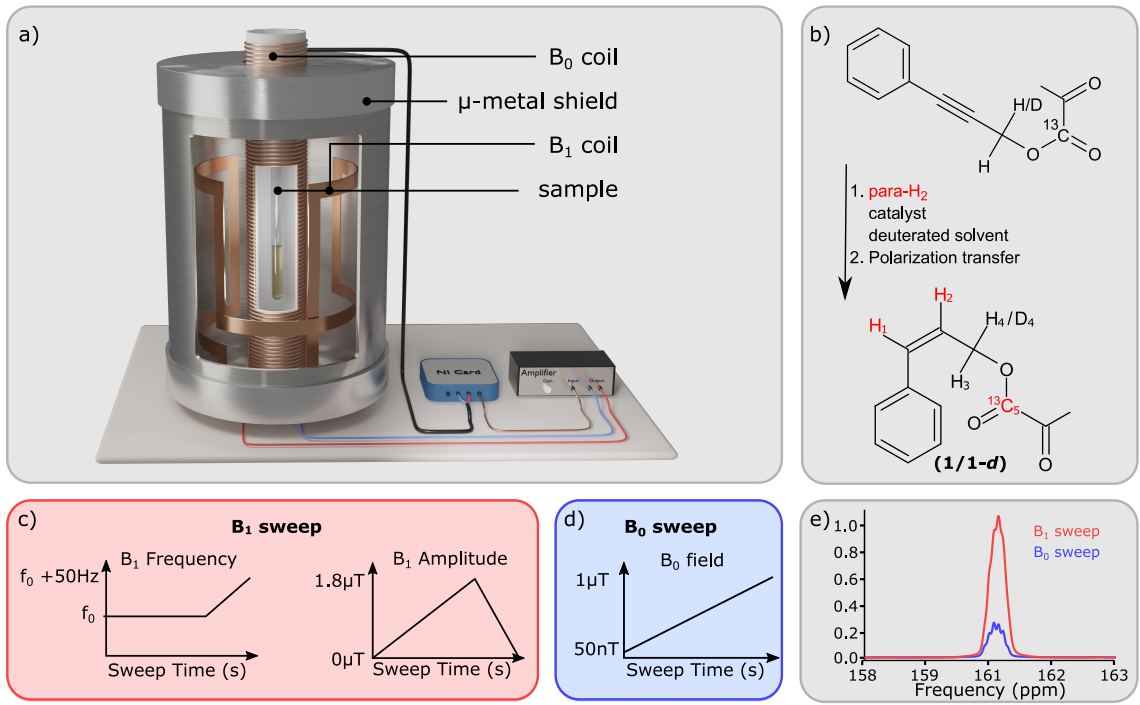}
    \caption{Panel (a) shows the experimental setup for the spin order transfer, including signal generator and amplifier. 
    (b) Shows the precursor materials and the reaction that yields molecules \textbf{1}/\textbf{1-\textit{d}}. 
    J-couplings of molecule \textbf{1} and \textbf{1-\textit{d}} are provided in the SI. 
    (c) Shows the frequency and amplitude profiles of the $B_1$ sweeps. 
    (d) Shows the amplitude profile of the $B_0$ sweep used 
    (e) Shows an exemplary resulting signal for molecule \textbf{1-\textit{d}} when applying both the $B_1$ sweep (orange) and the $B_0$ sweep (blue).
    }
    \label{fig:diamond and spectra}
\end{figure*}

Pyruvate is especially important as a probe because it can be used to study the metabolism of cancer and other pathologies \cite{Kurhanewicz2011, witneyKineticModelingHyperpolarized2011}. As pyruvate cannot be formed directly by means of a hydrogenation reaction, this molecule was for a long time inaccessible to hydrogenative PHIP, until 
the introduction of sidearm hydrogenation (PHIP-SAH) by Reineri and coworkers in 2015\cite{Reineri2015}.
In this technique, a pyruvate ester with an unsaturated bond in the side-arm is hydrogenated to introduce the entangled parahydrogen spins into the molecule. 
The non-equilibrium spin order is subsequently converted into ${}^{13}$C magnetization on pyruvate, after which the (hydrogenated) ester sidearm is removed via hydrolysis. 
This method has been demonstrated to be capable of producing pyruvate suitable for \emph{in vivo} imaging of metabolism in preclinical studies\cite{cavallari201813c}. For the PHIP-SAH method to be efficient, the transfer of spin-order from the parahydrogen molecules located on the side-arm to the adjacent ${}^{13}$C carboxyl group has to be optimized.

From a perspective of chemical stability and synthetic yield, propargylic pyruvates are more favourable than vinyl esters. However, in propargyl esters, mediating spins (typically the CH$_2$ $^1$H spins) are required to form a coupling network capable of relaying the polarization from the parahydrogen-derived $^1$Hs to the 1-${}^{13}$C spin of pyruvate within a reasonable amount of time (seconds).\cite{stewart2018long}

For hyperpolarization applications, it can be beneficial to introduce deuterium labels into molecules. 
Firstly, this can reduce the dipolar couplings experienced by other spins in the molecule as compared to $^1$H and thus can reduce relaxation rates. 
Secondly, as we show below, it can increase the efficiency of relaying polarization through the coupling network of a spin system generating higher ${}^{13}$C polarization of the metabolite. 
It is worth mentioning that the deuteration does not change the chemical properties such that, for example, hydrogenation efficiency and ester cleavage remain unaltered.
One major drawback of deuteration is ``polarization leakage,'' wherein any polarization transferred to a $^2$H spin is rapidly lost due to quadrupolar relaxation.
This is particularly problematic when using polarization-transfer schemes based on strong coupling at ultralow magnetic fields.\cite{TAYLER2019}
Consequently, the only options for such molecules have so far used coherence-transfer sequences at high magnetic fields (mT to T range) \cite{korchak2018over,goldmanDesignImplementation13C2006,schmidtLiquidstateCarbon13Hyperpolarization2017,haakeEfficientNMRPulse1996a}.

Here we present a general method of using radio-frequency (B$_1$) amplitude sweeps at low fields in the range of several $\mu$T to transfer polarization to ${}^{13}$C.
This field regime is particularly useful because the relaxation rates approach the extreme narrowing limit where $ T_1 = T_2 = T_{1\rho}$\cite{ivanov2008high,pfaff2017predicting}. 
Additionally, at low fields, the frequency difference between spins of the same isotope becomes negligible. 
It was this frequency splitting at high magnetic fields that in previous demonstrations limited $B_1$ sweeps to certain symmetric molecules   \cite{rodin2021constant,kozinenko2019polarization}.
We show that when combined with low fields, B$_1$ sweeps provide a robust method for polarization transfer to ${}^{13}$C, as demonstrated for a variety of molecules.
This is true even in the presence of quadrupolar relaxation sinks, which is shown to be particularly valuable because site-specific deuteration can be used to increase attainable polarization levels. 
%
%

\section{Results and Discussion}
In the radio-frequency (rf) amplitude sweeps introduced here ($B_1$ sweeps, see \cref{fig:diamond and spectra} (a,c)) the parahydrogen- $^1$H spin order is efficiently transferred to ${}^{13}$C polarization. %
We compare our scheme to a method based on ultra-low field magnetic field sweeps ($B_0$ sweeps). While we focus on two molecules, (\textit{Z})-cinnamyl pyruvate-1-${}^{13}$C and (\textit{Z})-cinnamyl-1-\textit{d} pyruvate-1-${}^{13}$C (\textbf{1} and \textbf{1-\textit{d}}, respectively, see \cref{fig:diamond and spectra}(b)), the analysis presented below and in the SI is given in sufficiently general form that it can be applied to other molecules having similar spin-network structures.

The $B_1$-sweep method consists of three main steps: (1) catalytic hydrogenation of the precursor molecule with para-enriched hydrogen gas in the presence of a magnetic field ($B_0 = \SI{100}{\micro\tesla}$); (2) adiabatically sweeping the amplitude of a radio-frequency field ($B_1$), resonant with the ${}^{13}$C Larmor frequency, from 0 to approximately \SI{1.8}{\micro\tesla} (in the rotating frame); and (3) adiabatically detuning the $B_1$ field's frequency while sweeping the amplitude back to 0. 

\begin{figure*}
    \centering
    \includegraphics[width=\textwidth]{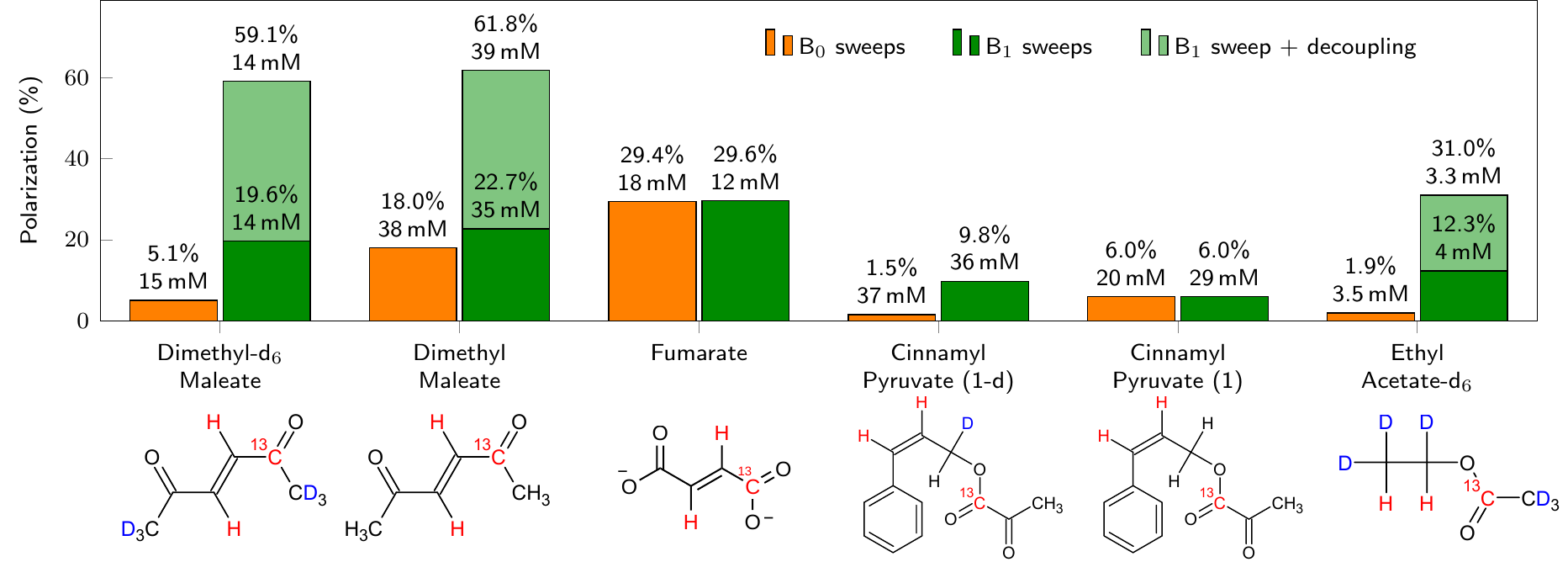}
    \caption{Comparison of $B_1$ sweeps and $B_0$ sweeps for six molecules. For all molecules studied $B_1$ sweeps perform comparable or significantly better than $B_0$ sweeps. Here, the hydrogenation is conducted in a water bath heated to \SI{60}{\celsius} (except for the case of fumarate, where it was heated to \SI{90}{\celsius}). The laboratory field during hydrogenation was approximately \SI{50}{\micro \tesla}. After a 10\,s hydrogenation, the samples were inserted into the magnetic shield, where either a $B_0$ or $B_1$ sweep was carried out immediately. For molecules where $^1$H decoupling was used, the hydrogenation was carried out at room temperature inside the magnetic shield (see experimental section for details). The concentrations were determined from $^1$H NMR measurements. 
    }
    \label{fig:comparison of molecules}
\end{figure*}

The main features of the system's evolution\footnote{The theoretical analysis and simulations do not consider relaxation, which is discussed later in the text.} through these steps are: (1) Generation of Parahydrogen-derived spin-order. 
Due to symmetry-breaking couplings to other spins in the molecule, the singlet state is not an eigenstate of the system.
Consequently, the initial state includes coherences which are averaged out due to the stochastic nature of the hydrogenation reaction\footnote{Some of these couplings may be suppressed by decoupling $^1$H spins from heteronuclei (see \cref{fig:comparison of molecules}) and is discussed below}. 
(2) Generation of ${}^{13}$C hyperpolarization that rotates with the rf-frequency in the x-y plane. 
(3) Movement of ${}^{13}$C polarization to the z direction for storage and transport.

The spin networks and J-couplings of molecules \textbf{1} and \textbf{1}-\textbf{\textit{d}} are given in \cref{fig:diamond and spectra} and in the SI. (b). The spin Hamiltonian for these systems in the absence of the radio-frequency field is ($\hbar=1$)
\begin{equation}\label{H}
    \mathcal{H} = \mathcal{H}_{B} 
    + \mathcal{H}^J_0 
    + \mathcal{H}^J_1,
\end{equation}
with
\begin{equation}\label{HB}
    \mathcal{H}_{B} = -2\pi \sum_{i} \gamma_i B_0 I_{iz},
\end{equation}
\begin{equation}\label{HJ0}
    \mathcal{H}^J_0 =
    2\pi\sum_{(i,j)}J_{ij}\textbf{I}_i \cdot \textbf{I}_j
    +
    2\pi\sum_{\langle i,j \rangle}J_{ij}I_{iz}I_{jz},
\end{equation}
and
\begin{equation}\label{HJ1}
    \mathcal{H}^J_1 = 2\pi\sum_{\langle i,j\rangle}J_{ij} \frac{1}{2}(I^+_i I^-_j + I^-_i I^+_j),
\end{equation}
where $(i,j) \And \langle i,j \rangle$ refer to summation over homonuclear and heteronuclear pairs, respectively. We define the single spin Zeeman operator states by $I_{z}\ket{\alpha}=+\frac{1}{2}\ket{\alpha}$ and $I_{z}\ket{\beta}=-\frac{1}{2}\ket{\beta}$ for spin $\frac{1}{2}$ nuclei and $I_{z}\ket{m}=m \ket{m}$ for $m=-1,0,1$ for spin $1$ nuclei. $\mathcal{H}$ is broken into three parts such that J-coupling terms are grouped into $\mathcal{H}_0^J$, which commutes with the magnetic field Hamiltonian $\mathcal{H}_B$, and $\mathcal{H}_1^J$ which does not.

For the (${}^{13}$C-resonant) $B_1$ sweeps we keep $B_0=\SI{100}{\micro\tesla}$ constant, and move to the interaction picture (i.e., rotating-frame) with respect to $\mathcal{H}_{B}$
and use the rotating-wave approximation to obtain:
\begin{equation}\label{Hrf}
    \mathcal{H}_{RF} = \mathcal{H}^J_0 
    - 2\pi \gamma_C B_1 I_{5x}
\end{equation}
where $B_1$ is the rf amplitude ($\gamma_C \And I_{5x}$ refer to the $^{13}$C, see \cref{fig:diamond and spectra} b)).

The spin network of molecule \textbf{1} has an important feature which limits the achievable ${}^{13}$C polarization — the symmetry of the \textit{mediating} geminal $^1$H spins (CH$_2$ group, labelled 3 \& 4).
Due to this symmetry, the mediating protons' eigenstates are the singlet and triplet states. 
In the mediating singlet state, all interactions between the parahydrogen $^1$Hs and the ${}^{13}$C nucleus are decoupled, resulting in a polarization limit of $75\%$.

It is important to note that this feature, namely, symmetric mediating hydrogens in CH$_2$ groups, does not depend on the values of the J-couplings and is not specific to this molecule. Similar limits hold for many spin networks that have one or more mediating CH$_2$ groups. We note that polarization limits due to symmetries of spin Hamiltonians have been discussed in previous works\cite{levitt2016symmetry,nielsen1995bounds}.

By replacing the mediating CH$_2$ group with a CHD group, the symmetry is broken and the aforementioned polarization limit is lifted. 
The standard $B_0$-sweep method, unfortunately, results in polarization of the fast-relaxing deuterium in this case, which is detrimental to ${}^{13}$C polarization. 
The $B_1$-sweep method we present here overcomes this issue by avoiding deuterium polarization, thereby allowing the exploitation of the “unlocked” remaining $25\%$ polarization, which is a 1.33 improvement factor over the $B_0$-sweep method for the undeuterated molecule. 
As discussed in the SI, by taking into account coherence loss during the hydrogenation stage the improvement factor increases to roughly 1.4, as the simulations and experimental results confirm.

\begin{figure}
    \centering
    \includegraphics[width=\columnwidth]{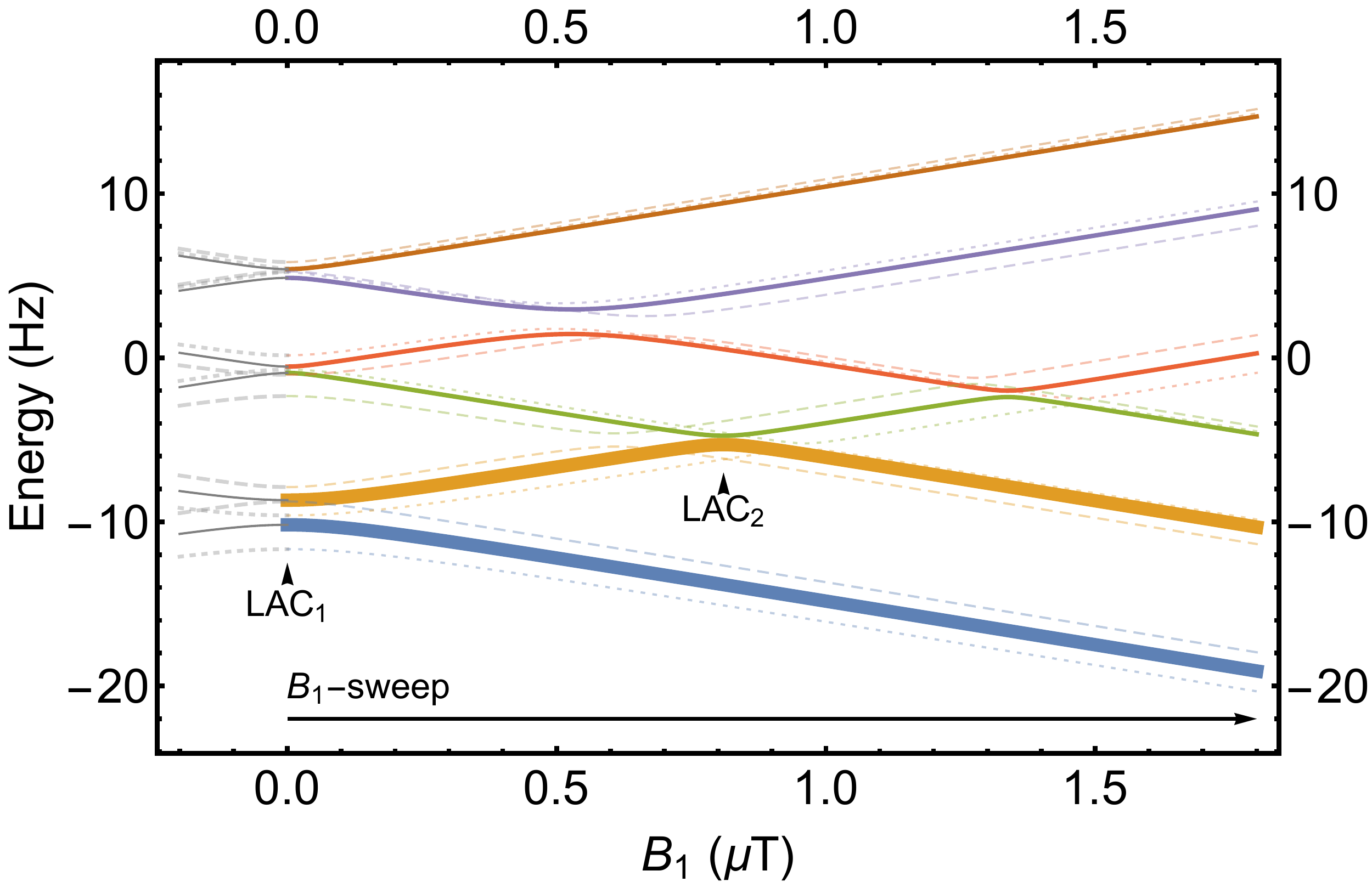}
    \caption{Eigenvalues of the Hamiltonian $\mathcal{H}_{RF}$ for molecule \textbf{1-\textit{d}}, as a function of the rf amplitude $B_1$. Only the eigenvalues relevant for the $B_1$-sweep method are shown (see text). Lines - eigenvalues for the subspace with $n=1$ and the deuterium state $m_s=0$, where $n$ is the eigenvalue of the magnetic quantum number operator and $m_s$ the eigenvalue of the deuterium's Zeeman operator $I_{4z}$ (see SI). Dashed (faint) - subspace with $n=1,m_s=1$. Dotted (faint) - subspace with $n=1,m_s=-1$. As can be seen, all three subspaces are qualitatively the same, thus, in the following we discuss only one of them ($n=1$ and $m_s=0$, lines). The two lowest eigenvalues (thick) are overpopulated after hydrogenation, at $B_1=0$. At high $B_1$, the three negative-slope eigenvalues correspond to states with ${}^{13}$C magnetization in the $x$ direction of the rotating frame. Sweeping the $B_1$ adiabatically to approximately \SI{1.8}{\micro\tesla} through the LACs results in eigenstates maintaining their population, allowing hyperpolarization of the ${}^{13}$C nucleus. 
    }
    \label{fig:LACsRFd1}
\end{figure}

We now describe the evolution of the system during the $B_1$ sweep. When adiabaticity is maintained, populations follow instantaneous eigenvalues\cite{kiryutin2015exploiting}. 
Immediately after hydrogenation, the highly populated states correspond predominantly to the lowest eigenvalues of $\mathcal{H}_{RF}|_{B_1=0}=\mathcal{H}_0^J$, a consequence of the size of $J_{12}$ (11.8\,Hz) compared to the other J-couplings involving $^1$H 1 \& 2 (see SI for further detail). 
Indeed, for sufficiently large $B_1$, the second term of $\mathcal{H}_{RF}$ dominates and the lowest eigenvalues correspond to ${}^{13}$C polarization along the $x$ direction in the rotating frame. 
We can thus conclude that the adiabatic $B_1$ sweep transfers the initial population imbalance into ${}^{13}$C polarization. 

For the system to evolve adiabatically during the $B_1$ sweep, $B_1(t)$ must be swept sufficiently slowly at the vicinity of the rotating-frame level anticrossings (LACs) involving the populated eigenvalues of $\mathcal{H}_{RF}$\cite{kiryutin2015exploiting}. The relevant LACs of the system can be best understood by examining the Hamiltonian's eigenvalues as a function of $B_1$\cite{eillsPolarizationTransferField2019,pravdivtsev2014highly} (see \cref{fig:LACsRFd1}). 
Unfortunately, simply plotting all levels for such large Hilbert spaces ($2^5=32$ for molecule \textbf{1} and $2^4*3=48$ for molecule \textbf{1-\textit{d}}) rarely results in something comprehensible (see SI Fig.\,S1).
We overcome this problem by decomposing the Hilbert space into interaction subspaces using the symmetries of the problem.

This reduces the complexity arising from the high dimensionality of the system, thus allowing easier identification of the relevant LACs, resonance conditions and involved states.


A detailed analysis, as well as a similar one for molecule \textbf{1}, and a discussion of the $^{13}$C spin dynamics during the sweep and step (3) of the method, are given in the SI.

The $B_0$-sweep method shares many of the features of the $B_1$-sweep method. It, too, consists of a hydrogenation stage, resulting in approximately the same density matrix. Here, a diabatic jump close to 0 magnetic field is performed \cite{Knecht2021} for the populated states to correspond to the lowest energy eigenstates at the beginning of the $B_0$ sweep\footnote{This is true far away from the LACs induced by $\mathcal{H}_1^J$.}. 
At high magnetic fields, the nucleus with the smallest gyromagnetic ratio is most effectively polarized (see SI for detailed discussion).
Therefore, while we expect little difference between $B_0$ sweeps and $B_1$ sweeps for molecule \textbf{1}, where the ${}^{13}$C spin is the small gyromagnetic nucleus, the method is expected to perform poorly for molecule \textbf{1-\textit{d}}, as polarization will be transferred to the deuterium which is also fast relaxing.

\begin{figure}
    \centering
    \includegraphics[scale=0.95]{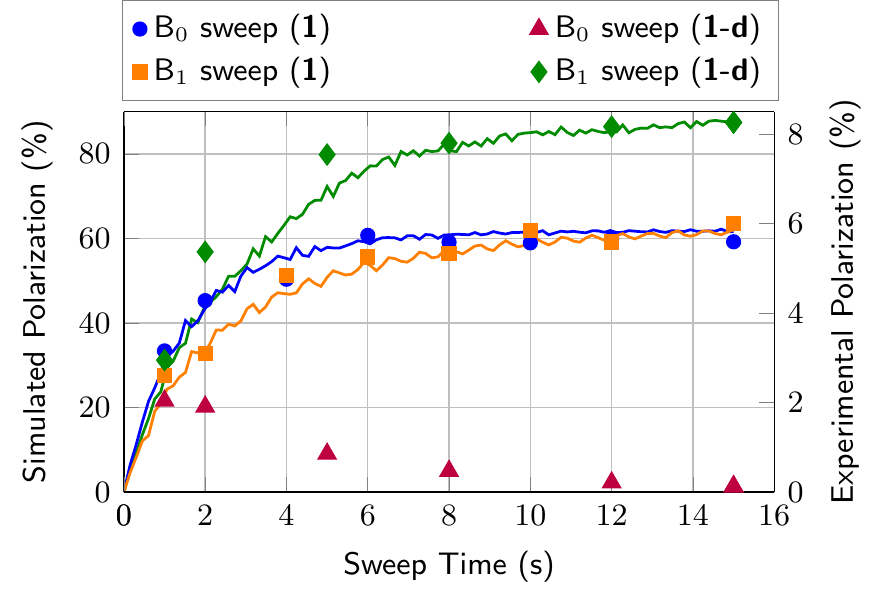}
    \caption{Hyperpolarization buildup kinetics using linear $B_1$ sweeps and $B_0$ sweeps. 
    The experimental data (symbols) is overlaid with spin dynamics simulations of the two transfer methods. 
    The orange squares and blue circles refer to polarization transfer in molecule \textbf{1} with CH$_2$ mediator spins connecting the carboxylic ${}^{13}$C spin with the parahydrogen spins. For this spin system, $B_1$ sweeps and $B_0$ sweeps perform comparably.
    The red triangles and green diamonds refer to polarization transfer in molecule \textbf{1-\textit{d}}, where one of the mediating spins is replaced with a deuteron. 
    Here, the polarization transfer efficiency is increased by approximately 40 percent when compared with molecule \textbf{1} (see SI for a detailed explanation). 
    Additionally, $B_0$ sweeps fail here because at longer sweep times polarization leakage through the deuterium spin dominates the dynamics. 
    When using $B_1$ sweeps, the $B_1$ amplitude was linearly increased from 0 to \SI{1.8}{\micro \tesla} and the hydrogenation was carried out in the $B_0$ field of \SI{100}{\micro \tesla}. 
    For the $B_0$ sweeps, the hydrogenation was performed at a static field of \SI{10}{\micro \tesla} generated inside the magnetic shield. 
    After hydrogenation was completed, the field was diabatically dropped to \SI{50}{\nano \tesla} and linearly swept up to \SI{1000}{\nano \tesla}.     }
    \label{fig:molecules and polarization}
\end{figure}

We apply $B_1$ and $B_0$ sweeps of varying durations to molecules \textbf{1} and \textbf{1-\textit{d}}. The shapes of the sweeps are depicted in \cref{fig:diamond and spectra} (c and d) and the dependence of the achieved polarization on the sweep duration for the two molecules is shown in \cref{fig:molecules and polarization}.
The points in the figure (circles, diamonds, triangles, and squares) are the measured data points from the experiment. The solid colored lines are numerical simulations (for more details, see SI) of the spin systems for both the $B_0$ and $B_1$ sweeps.  
For molecule \textbf{1}, the $B_0$ sweep (blue) performs well, rapidly rising to plateau at approximately 6\% polarization (60\% polarization transfer efficiency in simulation, as discussed in the SI) in the experiment. 
The simulations give excellent agreement.
Similar behavior, for molecule \textbf{1}, is observed for the $B_1$ sweep (orange) where polarization rises, albeit at a slightly slower rate, before plateauing at the same value. Again, simulations give excellent agreement.
It is expected that both methods perform equally well on molecule \textbf{1}.
Turning to molecule \textbf{1-\textit{d}}, the $B_0$ sweep (red triangles) performs exceptionally poorly. 
As the sweep progresses, almost no measurable polarization is left.
In contrast to this, the $B_1$ sweep reaches over 8\% polarization, exceeding both sweeps on molecule \textbf{1}. 
Simulations of this trajectory show excellent agreement once again. 

Relaxation of $^1$H and $^{13}$C spins during the sweeps does not appear to be an issue for the timescales we investigated, as the polarization level remains constant even for sweeps up to \SI{15}{\second} (see \cref{fig:molecules and polarization}).
However, for molecules containing a $^2$H spin, as the duration of the $B_0$ sweeps is increased, we see a severe deterioration in the polarization transfer efficiency. This is expected as the $^2$H on the molecule is polarized by the sweep, and this polarization is subsequently lost through relaxation.
When $B_1$ sweeps are used, no polarization is transferred to $^2$H and this relaxation pathway is effectively avoided. These effects combine to allow us to reach significantly higher polarization.
The theoretical efficiency of the sweeps on molecule \textbf{1-\textit{d}} is a factor of roughly \num{1.4} higher than for molecule \textbf{1}, as the singlet state of the CH$_2$ $^1$H spins which limits the efficiency of the transfer for the spin coupling network of molecule \textbf{1} is not present in molecule \textbf{1-\textit{d}}.
This is borne out in the experiment, where we see polarization reaching 8.25\% compared to 6.0\%.
The transfer speeds needed to ensure the transfer is adiabatic are similar to those of the $B_0$-field sweeps. It is important to note, that the partial deuteration does not increase the $T_1$ relaxation constant of the $^1$H spins introduced during hydrogenation (see SI). 
Thus, the increase in polarization does indeed come from the change of the spin coupling network topology and not from a decrease in relaxation losses during hydrogenation.

In addition, we show the generality of this method by comparing the $B_1$ sweeps and $B_0$ sweeps for numerous molecules, with and without deuteration. These experiments are shown in \cref{fig:comparison of molecules}. 
For some of these molecules, the parahydrogen-derived singlet state is not well protected by a dominant ${}^1$H--${}^1$H coupling. 
For these molecules, continuous-wave (CW) ${}^1$H decoupling was applied during the hydrogenation in order to avoid singlet-triplet mixing due to J-couplings between the parahydrogen ${}^1$H spins and either $^{13}$C or $^2$H nuclei of the molecule\cite{goldmanDesignImplementation13C2006,dagys2022deuteron,eriksson2022multiaxial}. In case of ethylacetate and dimethylmaleate this lead to significant increases in the achievable $^{13}$C polarization (see \cref{fig:comparison of molecules}).

\section{Conclusion}

In conclusion, we present a general method to produce highly spin-polarized molecules using PHIP even in the presence of fast-relaxing quadrupolar nuclei like deuterium. 
We demonstrate this method for a variety of molecules, including  interesting metabolic tracers such as pyruvate and fumarate. 
For all molecules tested, the level of polarization achieved with the $B_1$ sweeps was comparable to or higher than that achieved with the previously reported $B_0$ sweeps.


We expect that the discrepancy between the theoretically achievable polarization and the experimentally obtained one is largely due to relaxation losses during the hydrogenation process.
For molecules where short relaxation times impose a significant constraint on the spin-order transfer time, other transfer schemes using, e.g., constant-adiabaticity field sweeps\cite{rodin2019constant}, pulse sequences, SLIC \cite{barskiyNMRSLICSensing2016, dagysHyperpolarizationReadoutRapidly2022a} or optimal control\cite{khanejaTimeOptimalControl2001}, might be used. 
We stress that the presented decomposition of the Hamiltonian into sub-spaces is independent of the $B_1$ amplitude and thus can be helpful in the design of such alternative transfer schemes using both analytical and numerical methods.

For ((\textit{Z})-cinnamyl pyruvate, we demonstrate that selective $^2$H labeling, combined with $B_1$ sweeps allows a factor of \num{1.4} improvement to the final polarization in the experiment by breaking the symmetry of the mediating CH$_2$ $^1$Hs.


The experiments presented here are carried out at a $B_0$ field of \SI{100}{\micro \tesla}, but significantly lower fields are feasible. These fields, and field variances ($\leq$\,100\,nT), could also be reached without using a $\mu$-metal shield.



The polarization levels in this study are competitive with previously published work, and the relative ease of the technique in terms of instrumentation and tuning of parameters
%
%
%
offers the exciting potential to further develop PHIP as a competitive method 
for nuclear spin hyperpolarization.

\section*{Experimental Methods}

\subsubsection*{Hardware}
The low static magnetic field used in the described experiments was generated using a double layered solenoid piercing through a $\mu$-metal magnetic shield (MS-2, Twinleaf, USA) along the z-direction. This allows for the generation of the employed \SI{100}{\micro \tesla} field. 
The variance of the field was measured to be $\leq$\,100\,nT over the sample volume using a fluxgate magnetometer (Fluxmaster, Stefan Mayer Instruments, Germany).
The coil was powered using a common laboratory power supply, with the output attenuated over a \SI{1}{\kilo \ohm} high-power resistor. The rf excitation fields were generated using the built-in shim coils of the MS-2 shield perpendicular to the static $B_0$ field. The rf signals were generated by a digital to analogue converter (National Instruments 6363-USB) and amplified using an amplifier (KENMO M032S 12W Amplifier). The amplitude of the $B_1$ field was measured with the same fluxgate magnetometer (Fluxmaster).  

\subsubsection*{Sample Preparation}

For all experiments reported here, a reaction solution of aceton-d$_6$ containing \SI{4}{\milli \Molar} of Rh(dppb)(COD)BF$_4$ catalyst precursor was prepared. This solution was subsequently mixed with a solution of acetond-d$_6$ containing the starting material for the hydrogenation reaction, forming the desired PHIP molecules.
For the experiments yielding (\textit{Z})-cinnamyl pyruvate-1-${}^{13}$C or (\textit{Z})-cinnamyl-1-\textit{d} pyruvate-1-${}^{13}$C, \textbf{1-\textit{d}}, reported in \Cref{fig:molecules and polarization}, a solution of \SI{25}{\milli \Molar} of the 3-phenylpropargyl pyruvate 1-${}^{13}$C or 3-phenylpropargyl-1-d pyruvate 1-${}^{13}$C was prepared.
For the experiments yielding dimethyl maleate (\Cref{fig:comparison of molecules}), a concentration of \SI{50}{\milli \Molar} of dimethyl acetylenedicarboxylate (DMAD) or the fully deuterated and ${}^{13}$C labelled dimethyl-d$_6$ acetylenedicarboxylate-1-${}^{13}$C (DMAD-1-${}^{13}$C, d$_6$) was used.
For the production of hyperpolarized fumarate, a starting solution containing \SI{400}{\milli \Molar} of mono potassium acetylene dicarboylic acid together with \SI{8}{\milli \Molar} of the trans selective hydrogenation catalyst Ru(cp$^*$)MeCN$_3$ was prepared in D$_2$O according to a previously described procedure\cite{Knecht2021}.

\subsubsection*{Parahydrogen Bubbling Procedure}

Parahydrogen was enriched to 83\%-90\% in a liquid Helium cryostat by flowing Hydrogen gas over a catalyst.
For the experiments reported in \Cref{fig:molecules and polarization}, the samples are bubbled at room temperature under a hydrogen atmosphere of \SI{5}{\bar} for \SI{60}{\second} before a $B_1$ sweep or a $B_0$ sweep is applied. For the experiments reported in \cref{fig:comparison of molecules}, the hydrogenation was carried out in the laboratory magnetic field (approximately \SI{50}{\micro \tesla}) under elevated temperatures (\SI{55}{\celsius} for solutions using acetone-d$_6$ as the solvent and \SI{90}{\celsius} for D$_2$O as the solvent). For the experiments in \cref{fig:molecules and polarization} where vinyl acetate-d$_6$ and DMAD-1-$^{13}$C, d$_6$ were used as the precursors, the hydrogenation was carried out at room temperature inside the $\mu$-metal shield and $^1$H continuous-wave decoupling was performed with an amplitude of \SI{2}{\micro \tesla}.

\subsubsection*{Polarization Transfer Sweeps}

\paragraph*{$\mathbf{B_1}$ sweeps}

In our experiments, we used the adiabatic $B_1$ sweeps as depicted in \cref{fig:diamond and spectra}. In this scheme, the rf amplitude ($B_1$) is first swept from zero to approximately \SI{1.8}{\micro \tesla}, which assures that the populations of the spin system go through the relevant level avoided crossings described in the main text and SI.
The frequency of the $B_1$ field is kept on resonance with the ${}^{13}$C Larmor frequency at the given $B_0$ field (\SI{100}{\micro\tesla}). When not stated otherwise, the duration of this sweep was 7 seconds.
The result of the sweep (see \cref{fig:diamond and spectra}) is hyperpolarized ${}^{13}$C magnetization oscillating with the frequency of the RF field in the x-y plane.
To store the magnetization for transport, it is moved adiabatically to the z direction by linearly sweeping the frequency up by \SI{50}{\hertz}.
At the same time, the amplitude is reduced linearly to zero. This step is carried out within \SI{2}{\second}.
Many more schemes for transferring polarization using radio frequency drives are possible, and we discuss some of them in the conclusions to this article.

\paragraph*{$\mathbf{B_0}$ sweeps}

When $B_0$ sweeps were used, the sample was either kept at a static field of \SI{10}{\micro \tesla} (experiments reported in \cref{fig:molecules and polarization}) or inserted into the polarizer at that field (experiments reported in \cref{fig:comparison of molecules}). 
Subsequently, the field was diabatically dropped to \SI{50}{\nano \tesla} and linearly raised to \SI{1}{\micro \tesla}. Unless otherwise stated, the time of the transfer was 5 seconds.
The sample polarization was determined in the same manner as for the $B_1$ sweeps.

\subsubsection*{Acquisition}

Once the sweep has finished, the sample is carried to a benchtop NMR spectrometer (Fourier80, Bruker, USA) where the ${}^{13}$C NMR spectrum is acquired. 
To estimate the polarization, the $^1$H thermal signal is measured in a high-field spectrometer (Bruker \SI{400}{\mega \hertz}) and in the Fourier80 benchtop spectrometer. 
Subsequently, the sample concentration is calculated using an external reference. 
The polarization was determined by comparing the signal to the thermal signal of a reference sample (1-${}^{13}$C Methanol, 99.4 percent labeled). 
This approach was verified against acquiring the fully relaxed ${}^{13}$C spectrum of some of the highly concentrated ${}^{13}$C labeled molecules used in this study (for the remaining molecules this was not feasible within reasonable times due to the low SNR for ${}^{13}$C NMR measurements at the Fourier80 spectrometer).

\begin{acknowledgement}
A.M would like to acknowledge receiving funding from the European Union’s Horizon 2020 research and innovation programme under the Marie Sk\l{}odowska-Curie \mbox{QuSCo} (N$^\circ$ 765267), EU Quantum Flagship projects.
A.S. gratefully acknowledges the support of the Clore Israel Foundation Scholars Programme, the Israeli Council for Higher Education and the Milner Foundation.

\end{acknowledgement}

\begin{suppinfo}
The Supporting Information is available free of charge at [insert web address] and contains:
\begin{itemize}
\item J-coupling values.
\item Detailed theoretical considerations.
\item Numerical simulations of stability to $B_0$ variations.
\item Measurements of the lifetime of parahydrogen-derived $^1$H spin order.
\end{itemize}
\end{suppinfo}

\bibliography{references.bib}

\end{document}